\documentclass[11pt]{article}

\usepackage{graphicx,amsmath,amsfonts,amssymb, dcolumn,amsthm}
\usepackage{amsmath,amssymb,amsfonts}	
\usepackage{slashed}            
\usepackage{bbm}                
\usepackage{xspace}				

\numberwithin{equation}{section} 
\usepackage{tikz}
\usetikzlibrary{arrows,shapes}
\usetikzlibrary{trees}
\usetikzlibrary{matrix,arrows} 				
\usetikzlibrary{positioning}				
\usetikzlibrary{calc,through}				
\usetikzlibrary{decorations.pathreplacing}  
\usepackage{pgffor}							

\usetikzlibrary{decorations.pathmorphing}	
\usetikzlibrary{decorations.markings}
\tikzset{
    vector/.style={decorate, decoration={snake}, draw},
	provector/.style={decorate, decoration={snake,amplitude=2.5pt}, draw},
	antivector/.style={decorate, decoration={snake,amplitude=-2.5pt}, draw},
    fermion/.style={draw=black, postaction={decorate},
        decoration={markings,mark=at position .55 with {\arrow[draw=black]{latex}}}},
    fermionbar/.style={draw=black, postaction={decorate},
        decoration={markings,mark=at position .55 with {\arrow[draw=black]{latex reversed}}}},
    fermionnoarrow/.style={draw=black},
    gluon/.style={decorate, draw=black,
        decoration={coil,amplitude=4pt, segment length=5pt}},
    scalar/.style={dashed,draw=black, postaction={decorate},
        decoration={markings,mark=at position .55 with {\arrow[draw=black]{latex}}}},
    scalarbar/.style={dashed,draw=black, postaction={decorate},
        decoration={markings,mark=at position .55 with {\arrow[draw=black]{latex}}}},
    scalarnoarrow/.style={dashed,draw=black},
    electron/.style={draw=black, postaction={decorate},
        decoration={markings,mark=at position .55 with {\arrow[draw=black]{latex}}}},
	bigvector/.style={decorate, decoration={snake,amplitude=4pt}, draw},
}

\tikzstyle{block} = [draw, rectangle, 
    minimum height=3em, minimum width=6em]

\begin{document}

\title{\textbf{Gauge anomalies in Lorentz-violating QED}}
\author{ \textbf{Tiago R.~S.~Santos}\thanks{tiagoribeiro@if.uff.br}\  \ and \textbf{Rodrigo F.~Sobreiro}\thanks{sobreiro@if.uff.br}\\\\
\textit{{\small UFF $-$ Universidade Federal Fluminense,}}\\
\textit{{\small Instituto de F\'{\i}sica, Campus da Praia Vermelha,}}\\
\textit{{\small Avenida General Milton Tavares de Souza s/n, 24210-346,}}\\
\textit{{\small Niter\'oi, Rio de Janeiro, Brazil.}}}
\date{}
\maketitle

\begin{abstract}
In this work we study the issue of gauge anomalies in Lorentz-violating QED. To do so, we opt to use the Becchi-Rouet-Stora-Tyutin formalism within the algebraic renormalization approach, reducing our study to a cohomology problem. Since this approach is independent of the renormalization scheme, the results obtained here are expected to be general. We find that the Lorentz-violating QED is free of gauge anomalies to all orders in perturbation theory.
\end{abstract}

\section{Introduction}\label{intro}

Considerable studies have been done in the Standard Model Extension (SME) \cite{Colladay:1996iz,Colladay:1998fq,Kostelecky:2003fs}. This model looks for tiny deviations from Lorentz symmetry, which could be signs from underlying quantum-gravity theories \cite{Kostelecky:1988zi,Carroll:2001ws,Carmona:2002iv,GrootNibbelink:2004za,Alfaro:2004ur}. Specifically, in the minimal sector of SME (power-counting renormalizable) the studies are concentrated on Lorentz-violating electrodynamics, from phenomenological to theoretical points of view \cite{Bluhm:2005uj,Kostelecky:2008ts,Kostelecky:2000mm,Adam:2001ma}. Since this model is power-counting renormalizable, studies in its renormalizability properties and radiative corrections have been done \cite{Kostelecky:2001jc,deBerredoPeixoto:2006wz,DelCima:2012gb,Santos:2015koa}. For instance, in Ref.~\cite{Kostelecky:2001jc} the renormalizability is analyzed at one-loop order in perturbation theory. Applying dimensional regularization \cite{Bollini:1972bi,'tHooft:1972fi} in order to handle divergent Feynman integrals and using a mass-independent subtraction scheme, the authors are capable of showing that the Lorentz-violating electrodynamics is renormalizable at one-loop order. Remarkably, the model preserves gauge symmetry, i.e., the Ward identities are preserved at quantum level (one-loop). Moreover, this result is also confirmed through Pauli-Villars regularization \cite{Colladay:1998fq,Pauli:1949zm}. However, nothing is said about the gauge symmetry in finite terms. In fact, if one looks for radiative corrections, these two regularization methods show inequivalent results in the Chern-Simons like term generation (the so-called Chern-Simons generation controversy) \cite{Colladay:1998fq,Bonneau:2006ma,Chaichian:2000eh,Chen:2001yja,Chung:1998jv,Jackiw:1999yp,Jackiw:1999qq,PerezVictoria:2001ej,DelCima:2009ta,Scarpelli:2008fw,Brito:2007uc,Battistel:2000ms}. The analysis up to all orders in perturbation theory within the algebraic renormalization technique was discussed in \cite{DelCima:2012gb,Santos:2015koa}. In particular, in \cite{Santos:2015koa} it is formally shown that the Chern-Simons term does not renormalize and is not generated in radiative corrections.

Turning back to one-loop results, a step beyond Ref.~\cite{Kostelecky:2001jc} is presented in \cite{Franco:2013rp}, where the computation of the Feynman integrals related to the three-photon vertex diagrams is performed. It is shown that the contribution which comes from three-photon vertex diagrams is free of gauge anomalies at one-loop order, and it is conjectured that this property remains to all orders in perturbation theory. Essentially, they fixed the ambiguity of the internal momentum routing of the diagrams by requiring gauge invariance of the theory. They also stress the importance of this result for a complete proof of the all orders renormalizability of Lorentz-violating QED. In fact, once the anomaly parameter can be made to vanish at one-loop, a nonrenormalization theorem would be needed in order to guarantee this property up to all orders in perturbation theory. Still in this line, in \cite{Vieira:2015fra} are analyzed the one-loop contributions coming from vacuum-polarization and three-photon vertex diagrams. Besides divergent terms, the finite contributions from the mentioned diagrams are computed through the so-called implicit regularization, in order to circumvent the problem of the ambiguity in defining a $d$-dimensional $\gamma_5$-matrix algebra. Moreover, the momentum routing invariance is employed in order to ensure the gauge invariance of Lorentz-violating QED. Although the authors of \cite{Vieira:2015fra} disagree with the computation procedure employed in \cite{Franco:2013rp}, both works show which Ward identities of Lorentz-violating QED remain valid at one-loop order.

In this work we study the gauge anomaly issue in Lorentz-violating QED from the Becchi-Rouet-Stora-Tyutin (BRST) point of view within the algebraic renormalization approach \cite{Piguet:1995er}, avoiding, thus, possible problems from the regularization prescription. In this way, the analysis is reduced to a cohomology problem of the BRST operator. Essentially, we will employ the Symanzik method \cite{Symanzik:1969ek} of the external sources in order to control the Lorentz symmetry breaking and proceed with the BRST quantization of the model. Basically, we embed the Lorentz-violating QED in a more general theory, with well defined Lorentz and BRST symmetries. At the end of the study the theory is contracted down to the starting action. We emphasize that this method was vastly employed in Yang-Mills theories in order to control a soft BRST symmetry breaking in the Gribov-Zwanzinger scenario \cite{Zwanziger:1992qr,Dudal:2005na,Baulieu:2008fy,Baulieu:2009xr,Dudal:2011gd,Pereira:2013aza}, and also in the study of the renormalizability of Lorentz-violating Yang-Mills theory \cite{Santos:2014lfa}. We shall see that the Lorentz-violating QED is free of gauge anomalies to all orders in perturbation theory by verifying that the solution of the Slavnov-Taylor operator (functional form of the BRST operator) in the nontrivial sector of the cohomology is empty. This is verified through the extension of the Slavnov-Taylor operator to the quantum level, following the quantum action principle (QAP) prescription \cite{Piguet:1980nr}, and through the algebra obeyed by the Slavnov-Taylor operator and a set of extra Ward identities (which are also anomalies free).

Let us clarify a few points about anomalies. An anomaly is, essentially, the breaking of a Ward identity at quantum level. One speaks of gauge anomaly if the Slavnov-Taylor identity breaks down. It worthwhile to emphasize that relevant anomalies are independent of any regularization prescription employed in the computation of the Feynman integrals. In fact, relevant anomalies are usually related to observable physical phenomena \cite{Bell:1969ts,Adler:1969gk,Bardeen:1969md}. Moreover, if one uses cohomology properties in order to characterize the anomalies present in the theory, the relevant anomalies correspond to a nonempty solution of the nontrivial sector of the cohomology. Every time this happens, a nonrenormalization theorem is needed in order to establish the absence (or not) of quantum corrections on the anomaly parameter at high-orders in perturbation theory. Because of the importance of the gauge symmetry in renormalizability and unitarity analysis of gauge theory \cite{Becchi:1975nq,Tyutin:1975qk,Kugo:1979gm,Gross:1972pv}, usually one looks for some mechanism yielding a vanishing anomaly parameter, e.g., sum of all fermions in a family. In the case that this is possible, a nonrenormalization theorem comes in to guarantee this property to all orders. If this is possible the theory is said to be anomaly free. On the other hand, when the anomaly depends on the regularization scheme, one characterizes an irrelevant anomaly. In the cohomology point of view, the irrelevant anomalies belong to the trivial sector of the cohomology. Thus, besides the restoration of Lorentz symmetry, to employ the BRST techniques is a safe way to study the anomalies of the Lorentz-violating QED.

This work is organized as follows: In Sec.~\ref{QED} we present the Lorentz-violating electrodynamics, with our conventions, its properties, and its definitions. In Sec.~\ref{QUATIZATION}, the BRST quantization of the model with the extra set of auxiliary sources is provided. In Sec.~\ref{Anomalies}, we study the possible anomalies of the model by studying a cohomology problem. Final comments are displayed in Sec.~\ref{FINAL}.

\section{Lorentz-violating electrodynamics}\label{QED}

QED is a gauge theory based on the $U(1)$ symmetry group describing the interaction between the electromagnetic and the Dirac fields. The minimal sector of the QED extension consists in an extra sector that breaks the Lorentz symmetry in such a way that the gauge symmetry is still preserved. The Lorentz symmetry violation manifests through couplings between background external fields and composite operators with the dimension bounded by four (power-counting renormalizability). The action for this model reads 
\begin{eqnarray}
S_{QEDex}&=&S_{QED}+S_{LV}\;,
\label{0}
\end{eqnarray}
where
\begin{eqnarray}
S_{QED}&=&\int d^4x\left[-\frac{1}{4}F^{\mu\nu}F_{\mu\nu}+\overline{\psi}(i\gamma^{\mu}D_{\mu}-m)\psi\right]\;
\label{1}
\end{eqnarray}
is the classical action of the usual QED. The covariant derivative is defined as $D_{\mu}\equiv\partial_{\mu}+ieA_{\mu}$, the field strength is written as $F_{\mu\nu}\equiv\partial_{\mu}A_{\nu}-\partial_{\nu}A_{\mu}$ and $A_\mu$ is the gauge field. The parameter $m$ is the electron mass and $e$ the electromagnetic coupling constant. The second term in \eqref{0} is the Lorentz-violating action,
\begin{eqnarray}
S_{LV}&=&\int d^4x\left[\epsilon_{\mu\nu\alpha\beta}v^{\mu}A^{\nu}\partial^{\alpha}A^{\beta}-\frac{1}{4}\kappa_{\alpha\beta\mu\nu}F^{\alpha\beta}F^{\mu\nu}+\overline{\psi}i\Gamma^{\mu}D_{\mu}\psi-\overline{\psi}M\psi\right]\;,
\label{2}
\end{eqnarray}
where
\begin{eqnarray}
\Gamma^{\mu}&\equiv&c^{\nu\mu}\gamma_{\nu}+d^{\nu\mu}\gamma_5\gamma_{\nu}+e^{\mu}+if^{\mu}\gamma_5+\frac{1}{2}g^{\alpha\beta\mu}\sigma_{\alpha\beta}\ ,\nonumber\\
M&\equiv&im_5\gamma_5+a^{\mu}\gamma_{\mu}+b^{\mu}\gamma_5\gamma_{\mu}+\frac{1}{2}h^{\mu\nu}\sigma_{\mu\nu}\;.
\label{3}
\end{eqnarray}
The Lorentz symmetry violation in the fermionic sector manifests itself through the following constant tensorial fields: $c^{\nu\mu}$, $d^{\nu\mu}$, $e^{\mu}$, $f^{\mu}$, $g^{\alpha\beta\mu}$, $m_5$, $a^{\mu}$, $b^{\mu}$, and $h^{\mu\nu}$. They define privileged directions in spacetime and are constant tensors. Couplings that take into account tensorial fields with an even number of indices preserve the charge-parity-time (CPT) discrete mapping, while those that include an odd number of indices do not preserve CPT. The tensorial fields $c^{\nu\mu}$, $d^{\nu\mu}$, $e^{\mu}$, $f^{\mu}$, and $g^{\alpha\beta\mu}$ are dimensionless while $m_5$, $a^{\mu}$, $b^{\mu}$, and $h^{\mu\nu}$ have mass dimension 1. The tensorial fields $c^{\nu\mu}$ and $d^{\nu\mu}$ are taken to be traceless. The tensor field $h^{\mu\nu}$ is antisymmetric, and $g^{\alpha\beta\mu}$ is antisymmetric only on its first two indices. At the gauge field sector, the Lorentz violation is characterized by the field $v^{\mu}$, with mass dimension 1, and $\kappa_{\alpha\beta\mu\nu}$, which is dimensionless. This tensor obeys the same properties of the Riemann tensor and is double traceless, 
\begin{eqnarray}
\kappa_{\alpha\beta\mu\nu}\;=\;\kappa_{\mu\nu\alpha\beta}&=&-\kappa_{\beta\alpha\mu\nu}\ ,\nonumber\\
\kappa_{\alpha\beta\mu\nu}+\kappa_{\alpha\mu\nu\beta}+\kappa_{\alpha\nu\beta\mu}&=&0\ ,\nonumber\\
\kappa^{\mu\nu}_{\phantom{\mu\nu}\mu\nu}&=&0\ .
\label{3aaa}
\end{eqnarray}
Although the action \eqref{0} violates discrete symmetries, these symmetries still have an important role. In fact, the quantum corrections on the classical action \eqref{0} are limited by properties of the background tensors. This means that, for instance, combinations between background tensor fields and/or mass parameters that lead to a pseudovector are avoided to couple with a vectorial composite operator. This property, of course, is not expected to be violated at quantum level. In fact, one-loop explicit computations confirm this property \cite{Kostelecky:2001jc,Santos:2015koa}. Otherwise, there would be no distinction between vectorial and pseudovectorial background fields, for instance.

\section{BRST quantization of Lorentz-violating electrodynamics}\label{QUATIZATION}

As is widely known, the Faddeev-Popov fields decouple from the dynamics because they do not interact with the gauge field. Thus, in principle, they could be neglected. Indeed, in the electrodynamics in the Landau gauge the ghost fields are irrelevant; but they are relevant if one is interested in looking to QED from the BRST cohomology point of view. In fact, since the BRST operator is nilpotent, the solutions of the quantum theory are restricted to a cohomology problem. This property is very useful in the study of the renormalizability, anomalies, and unitarity of gauge theories. In particular, renormalizability and anomalies are studied by seeking the most general solutions in the space of the local integrated polynomial in the fields and their derivatives with zero and one ghost numbers, respectively. Another useful feature of the BRST approach is that the introduction of the fields $b,\;\bar{c}$, and $c$ provides a set of extra Ward identities, which are very useful in the understanding of the nonrenormalizabity of certain operators \cite{Santos:2015koa}. Hence, we will employ the BRST quantization method in this entire work.

The BRST quantization allows one to fix the gauge and deal with the loss of the gauge symmetry in a systematic way. In fact, by introducing the Lautrup-Nakanishi field $b$ (a Langrange multiplier) and the Faddeev-Popov ghost and antighost fields $c$ and $\overline{c}$ and defining the following BRST transformations acting on the fundamental fields:
\begin{eqnarray}
sA_{\mu}&=&-\partial_{\mu}c\;,\nonumber\\
sc&=&0\;,\nonumber\\
s\psi&=&iec\psi\;,\nonumber\\
s\overline{\psi}&=&ie\overline{\psi}c\;,\nonumber\\
s\overline{c}&=&b\;,\nonumber\\
sb&=&0\;,\label{4a}
\end{eqnarray}
where $s$ is the nilpotent BRST operator, it is possible to treat the quantization of electrodynamics with an off-shell BRST symmetry. Choosing the Landau gauge $\partial_{\mu} A^{\mu}=0$, for simplicity, the gauge fixed action reads
\begin{eqnarray}
S&=&S_{QED}+S_{gf}\;,
\label{4b}
\end{eqnarray}
where 
\begin{eqnarray}
S_{gf}&=&s\int d^4x\overline{c}\partial_{\mu}A^{\mu}=\int d^4x\left(b\partial_{\mu}A^{\mu}+\overline{c}\partial^2c\right)
\label{6}
\end{eqnarray}
is the gauge fixing action enforcing the Landau gauge condition. In the sequel we display the quantum numbers of the fields and background tensors in Tables \ref{table1} and \ref{table2}, respectively.
\begin{table}[h]
\centering
\begin{tabular}{|c|c|c|c|c|c|c|}
	\hline 
Fields & $A$ & $b$ &$c$ & $\overline{c}$ & $\psi$ & $\overline{\psi}$ \\
	\hline 
UV dimension & $1$ & $2$ & $0$ & $2$ & $3/2$& $3/2$\\ 
Ghost number & $0$ & $0$ & $1$& $-1$& $0$& $0$\\ 
Spinor number & $0$ & $0$ & $0$& $0$& $1$&$-1 $\\ 
Statistics & $0$& $0$ & $1$ & $-1$& $1$ & $-1$ \\ 
\hline 
\end{tabular}
\caption{Quantum numbers of the fields.}
\label{table1}
\end{table}
\begin{table}[h]
\centering
\begin{tabular}{|c|c|c|c|c|c|c|c|c|c|c|c|}
	\hline 
Tensors & $v$ & $\kappa$ & $c$ & $d$ & $e$ & $f$ & $g$ & $m_5$ & $a$ & $b$ & $h$ \\
	\hline 
UV dimension & $1$ & $0$ & $0$ & $0$ & $0$ & $0$ & $0$ & $1$ & $1$ & $1$ & $1$ \\ 
Ghost number & $0$ & $0$ & $0$ & $0$ & $0$ & $0$ & $0$ & $0$ & $0$ & $0$ & $0$ \\ 
Spinor number & $0$ & $0$ & $0$ & $0$ & $0$ & $0$ & $0$ & $0$ & $0$ & $0$ & $0$  \\ 
Statistics & $0$ & $0$  & $0$ & $0$  & $0$ & $0$ & $0$ & $0$ & $0$ & $0$ & $0$ \\ 
\hline 
\end{tabular}
\caption{Quantum numbers of the background tensors.}
\label{table2}
\end{table}

To deal with the Lorentz-violating sector within the BRST quantization, extra care is demanded. In fact, the BRST quantization requires that all objects coupled to the BRST invariant operators must also be BRST invariant; on the other hand, objects coupled to the BRST noninvariant operators must have a BRST counterpart, i.e., they must belong to a BRST doublet, in order to guarantee the BRST symmetry. For instance, the background tensor $v^\mu$ is coupled to a noninvariant BRST operator (but it is on-shell gauge invariant in the action -- it is assumed that surface terms can be safely discarded and $\partial_{[\mu}v_{\nu]}=0$ must be employed). Nevertheless, dropping out surface terms requires that the integrand is composed by smooth functions at the compact support; however, nothing can be said about the background tensors. Moreover, it is not always true that a classical on-shell symmetry is preserved at quantum level. In fact, the QAP, which is the cornerstone of the algebraic renormalization, is applicable to local, power-counting renormalizable, and Lorentz invariant theories. Basically the QAP establishes that the extension of the vertex functional must be a local polynomial respecting the power-counting criteria. Within the Green function formalism, the proofs of the theorems on which the QAP is based assume Lorentz covariance of the Green functions \cite{Lowenstein:1971vf, Clark:1976ym, Lam:1972mb}. Also there exists a proof of the QAP without using Green functions, but within a well-defined causal structure \cite{Duetsch:2000nh}. 

Here, there are two distinct types of Lorentz transformations -- the observer and particle Lorentz transformations -- with the loss of the Lorentz covariance under particle transformation. To handle these issues, we will use the Symanzik external sources method \cite{Symanzik:1969ek}. This method consists in introducing a set of external fields in order to control broken symmetries. Here, this means that we will embed the Lorentz-violating action into a more general theory that respects Lorentz and CPT symmetry, and also BRST symmetry. We will proceed with this by treating each of the background tensors as an external classical source. However, there are two distinct situations due to the classes of composite operators. The sources that are coupled with BRST invariant composite operators also must be BRST invariant. On the other hand, sources coupled with BRST noninvariant composite operators must belong to a BRST doublet with their BRST counterpart, in order to ensure the BRST symmetry. Since the CPT-even bosonic violating term and all fermionic breaking terms are BRST invariant, they will couple to invariant sources. Thus, we define the following set of invariant sources:
\begin{eqnarray}
s\bar{\kappa}_{\alpha\beta\mu\nu}=sC^{\nu\mu}=sD^{\nu\mu}=sE^{\mu}=sF^{\mu}=sG^{\alpha\beta\mu}=sM_5=s\bar{A}^{\mu}=sB^{\mu}=sH^{\mu\nu}=0\;.
\label{10}
\end{eqnarray}
On the other hand, the CPT-odd bosonic violating term will be coupled to a BRST doublet, 
\begin{eqnarray}
s\lambda_{\mu\nu\alpha}&=&J_{\mu\nu\alpha}\;, \nonumber\\
sJ_{\mu\nu\alpha}&=&0\;.
\label{9}
\end{eqnarray}
The quantum numbers of the sources are displayed in Table~\ref{table3}. Eventually, to reobtain the starting action \eqref{4b}, these sources will attain the following physical values:
\begin{eqnarray}
J_{\mu\nu\alpha}\mid_{phys}&=&v^{\beta}\epsilon_{\beta\mu\nu\alpha}\;,\nonumber\\
\lambda_{\mu\nu\alpha}\mid_{phys}&=&0\ ,\nonumber\\
\bar{\kappa}_{\alpha\beta\mu\nu}\mid_{phys}&=&\kappa_{\alpha\beta\mu\nu}\;,\nonumber\\
C^{\nu\mu}\mid_{phys}&=&c^{\nu\mu}\;,\nonumber\\
D^{\nu\mu}\mid_{phys}&=&d^{\nu\mu}\;,\nonumber\\
E^{\mu}\mid_{phys}&=&e^{\mu}\;,\nonumber\\
F^{\mu}\mid_{phys}&=&f^{\mu}\;,\nonumber\\
G^{\alpha\beta\mu}\mid_{phys}&=&g^{\alpha\beta\mu}\;,\nonumber\\
M_5\mid_{phys}&=&m_5\;,\nonumber\\
\bar{A}^{\mu}\mid_{phys}&=&a^{\mu}\ ,\nonumber\\
B^{\mu}\mid_{phys}&=&b^{\mu}\;,\nonumber\\
H^{\mu\nu}\mid_{phys}&=&h^{\mu\nu}\;.
\label{12}
\end{eqnarray}
\begin{table}[h]
\centering
\begin{tabular}{|c|c|c|c|c|c|c|c|c|c|c|c|c|c|c|}
	\hline 
Sources & $Y$ & $\overline{Y}$ &$\lambda$ & $J$ & $\bar{\kappa}$ & $C$ & $D$ & $E$ & $F$ & $G$ & $M_5$ & $\bar{A}$ & $B$ & $H$ \\
	\hline 
UV dimension & $5/2$ & $5/2$ & $1$ & $1$ & $0$ & $0$ & $0$ & $0$ & $0$ & $0$ & $1$ & $1$ & $1$ & $1$ \\ 
Ghost number & $-1$ & $-1$ & $-1$ & $0$ & $0$ & $0$ & $0$ & $0$ & $0$ & $0$ & $0$ & $0$ & $0$ & $0$ \\ 
Spinor number & $1$ & $-1$ & $0$ & $0$ & $0$ & $0$ & $0$ & $0$ & $0$ & $0$ & $0$ & $0$ & $0$ & $0$  \\ 
Statistics &  $0$ & $-2$ & $-1$ & $0$ & $0$  & $0$ & $0$  & $0$ & $0$ & $0$ & $0$ & $0$ & $0$ & $0$ \\ 
\hline 
\end{tabular}
\caption{Quantum numbers of the sources.}
\label{table3}
\end{table}
The embedding of the bosonic sector of Lorentz-violating QED reads
\begin{eqnarray}
S_{B}&=&s\int d^4x\;\lambda_{\mu\nu\alpha}A^{\mu}\partial^{\nu}A^{\alpha}-\frac{1}{4}\int d^4x\;\bar{\kappa}_{\alpha\beta\mu\nu}F^{\alpha\beta}F^{\mu\nu}\;\nonumber\\
&=&\int d^4x\left(J_{\mu\nu\alpha}A^{\mu}\partial^{\nu}A^{\alpha}+\lambda_{\mu\nu\alpha}\partial^{\mu}c\partial^{\nu}A^{\alpha}\right)-\frac{1}{4}\int d^4x\;\bar{\kappa}_{\alpha\beta\mu\nu}F^{\alpha\beta}F^{\mu\nu}\;,
\label{7}
\end{eqnarray}
while the embedding of the Lorentz-violating term for the fermionic sector is given by
\begin{eqnarray}
S_{F}&=&\int d^4x\left[i\left(C^{\nu\mu}\overline{\psi}\gamma_{\nu}D_{\mu}\psi+D^{\nu\mu}\overline{\psi}\gamma_{5}\gamma_{\nu}D_{\mu}\psi+E^{\mu}\overline{\psi}D_{\mu}\psi+iF^{\mu}\overline{\psi}\gamma_{5}D_{\mu}\psi\right.\right.\nonumber\\
&+&\left.\left.\frac{1}{2}G^{\alpha\beta\mu}\overline{\psi}\sigma_{\alpha\beta}D_{\mu}\psi\right)-\left(iM_5\overline{\psi}\gamma_5\psi+\bar{A}^{\mu}\overline{\psi}\gamma_{\mu}\psi+B^{\mu}\overline{\psi}\gamma_5\gamma_{\mu}\psi+\frac{1}{2}H^{\mu\nu}\overline{\psi}\sigma_{\mu\nu}\psi\right)\right]\;.\nonumber\\
\end{eqnarray}
Moreover, to control the nonlinear BRST transformations of the original fields, we need one last set of external BRST invariant sources, namely $\overline{Y}$ and $Y$, to introduce the following action:
\begin{eqnarray}
S_{ext}&=&\int d^4x\left(\overline{Y}s\psi-s\overline{\psi}Y\right)\;=\;\int d^4x\left(ie\overline{Y}c\psi-ie\overline{\psi}cY\right)\;.
\label{13}
\end{eqnarray}
Thus, the most complete BRST invariant action is given by
\begin{eqnarray}
\Sigma&=&S+S_B+S_F+S_{ext}\:.
\label{15}
\end{eqnarray}
Indeed, this is not the most general BRST invariant action; it is easy to note that extra combinations among sources are possible -- vacuum terms, including the electron mass. However, this will not be important for the study of gauge anomalies. They are important, however, in the stability study of the model; see \cite{Santos:2015koa}. The quantum numbers of the sources follow the quantum numbers of the background fields, as displayed in Table \ref{table3}. It is easy to check that the action $\Sigma$, \eqref{15}, at the physical value of the sources \eqref{12}, reduces to
\begin{equation}
\Sigma_{phys}=S_{QED}+S_{LV}+S_{gf}\;.
\label{15aab}
\end{equation}

It is worthwhile to note that, once we have introduced the Symanzik sources, the action displayed in \eqref{15} does not correspond to the actual physical action anymore. The Lorentz-violating QED action was embedded into a more general action. In fact, the embedded action enjoys Lorentz, CPT, and BRST symmetries, and it will be the action submitted to the perturbative treatment within the algebraic renormalization approach. Moreover, the CPT-odd Lorentz-violating sector of the gauge field was embedded with the help of a BRST doublet, namely $J$ and $\lambda$. From the general results of cohomology \cite{Piguet:1995er}, this sector does not belong to the physical sector of the cohomology, just like the gauge fixing action. In fact, as aforementioned, in this stage the theory does not correspond to the physical theory. Only after the study of the gauge anomalies is the larger theory contracted down to the original action \eqref{15aab}. Furthermore, when the physical values of the sources \eqref{12} are taken, the BRST symmetry is explicitly broken and the CPT-odd Lorentz-violating term of the gauge field is ``released.''\footnote{Although the Symanzik method is not a regularization method, it can be compared with by analogy. Let us consider, for instance, dimensional regularization \cite{Bollini:1972bi,'tHooft:1972fi}. In dimensional regularization, a Feynman integral is embedded in a bigger space with complex dimension, say, $d=4-\epsilon$, containing spacetime. Only after the divergences are reabsorbed is the limit to four dimensions taken. One of the advantages of this method for gauge theories is that the gauge symmetry is preserved in the regularization process, avoiding problems for unitarity of the theory. Nevertheless, one does not speak about unitarity in the regularized phase of the theory. Moreover, in dimensional regularization an arbitrary mass parameter $\mu$ is introduced in order to keep the coupling constant dimensionless. One of the consequences of this is that the theory will be independent from the renormalization point, leading to the renormalization group flow. In the Symanzik method, effects from the procedure of the broken symmetries' control can also appear at the end, a kind of relic of the most general phase of the theory: In the present case, when the external sources attain their physical values (see \cite{Santos:2014lfa}), vacuum terms are generated.} At this stage, one might study the effects of the CPT-odd Lorentz-violating sector on the unitarity of the action \eqref{15aab} \cite{Adam:2001ma,BaetaScarpelli:2003yd}. 

\section{Gauge anomalies in Lorentz-violating QED}\label{Anomalies}

To proceed with the algebraic analysis on the possible gauge anomalies in Lorentz-violating QED, we will display the most complete set of Ward identities enjoyed by the classical action \eqref{15}, and in the sequel we shall study their extension to quantum level.

\subsection{Ward identities}\label{WI}

The action \eqref{15} displays a rich set of Ward identities, listed below.

\begin{itemize}
	\item Slavnov-Taylor identity
	\begin{eqnarray}
\mathcal{S}(\Sigma)&=&\int d^4x\left(-\partial_{\mu} c\frac{\delta \Sigma}{\delta A_{\mu}}+\frac{\delta \Sigma}{\delta \overline{Y}}\frac{\delta \Sigma}{\delta \psi}-\frac{\delta \Sigma}{\delta Y}\frac{\delta \Sigma}{\delta \overline{\psi}}+b\frac{\delta \Sigma}{\delta \overline{c}}+J_{\mu\nu\alpha}\frac{\delta \Sigma}{\delta\lambda_{\mu\nu\alpha}}\right)=0\;.
\label{16}
\end{eqnarray}

\item Gauge fixing and antighost equations
\begin{eqnarray}
\frac{\delta \Sigma}{\delta b}&=&\partial_{\mu}A^{\mu}\;,\nonumber\\
\frac{\delta \Sigma}{\delta \overline{c}}&=&\partial^2c\;.
\label{17}
\end{eqnarray}

\item Ghost equation
\begin{eqnarray}
\frac{\delta \Sigma}{\delta c}&=&\Delta_{cl}\;,
\label{18}
\end{eqnarray}
where
\begin{eqnarray}
\Delta_{cl}&=&\partial^\mu\left(\lambda_{\mu\nu\alpha}\partial^\nu A^\alpha\right)-\partial^2\overline{c}+ie\overline{Y}\psi+ie\overline{\psi}Y\;.
\label{18b}
\end{eqnarray}
\item Rigid symmetry
\begin{eqnarray}
\mathcal{W}_{rig}(\Sigma)&=&\int d^4x\left(e\frac{\delta \Sigma}{\delta \psi}\psi+e\overline{Y}\frac{\delta \Sigma}{\delta \overline{Y}}+e\overline{\psi}\frac{\delta \Sigma}{\delta \overline{\psi}}-e\frac{\delta \Sigma}{\delta Y}Y\right)=0\;.
\label{19}
\end{eqnarray}
\end{itemize}
In Eqs.~\eqref{17} and \eqref{18}, the breaking terms are linear in the fields. Thus, they will remain at classical level \cite{Piguet:1995er}.

Let $\mathcal{F}$ be a general functional with an even ghost number. The Slavnov-Taylor operator acting on $\mathcal{F}$ is defined as
\begin{eqnarray}
\mathcal{S}(\mathcal{F})&=&\int d^4x\left(-\partial_{\mu} c\frac{\delta \mathcal{F}}{\delta A_{\mu}}+\frac{\delta \mathcal{F}}{\delta \overline{Y}}\frac{\delta \mathcal{F}}{\delta \psi}-\frac{\delta \mathcal{F}}{\delta Y}\frac{\delta \mathcal{F}}{\delta \overline{\psi}}+b\frac{\delta \mathcal{F}}{\delta \overline{c}}+J_{\mu\nu\alpha}\frac{\delta \mathcal{F}}{\delta\lambda_{\mu\nu\alpha}}\right)\;.
\label{20}
\end{eqnarray}
Moreover, we can define the linearized Slavnov-Taylor operator, emerging from a perturbative expansion of \eqref{20} with $\mathcal{F}$ being the leading term,
\begin{eqnarray}
\mathcal{S}_{\mathcal{F}}=\int d^4x\left(-\partial_{\mu} c\frac{\delta}{\delta A_{\mu}}+\frac{\delta \mathcal{F}}{\delta \overline{Y}}\frac{\delta }{\delta \psi}+\frac{\delta \mathcal{F}}{\delta\psi}\frac{\delta}{\delta\overline{Y}}-\frac{\delta \mathcal{F}}{\delta Y}\frac{\delta}{\delta \overline{\psi}}-\frac{\delta \mathcal{F}}{\delta\overline{\psi}}\frac{\delta}{\delta Y}+b\frac{\delta}{\delta \overline{c}}+J_{\mu\nu\alpha}\frac{\delta }{\delta\lambda_{\mu\nu\alpha}}\right)\;.
\label{21} 
\end{eqnarray}
Also, the following identities are satisfied:
\begin{eqnarray}
\mathcal{S}_{\mathcal{F}}\mathcal{S}(\mathcal{F})&=&0\;,\;\;\forall\;\;\mathcal{F}\;,\nonumber\\
\mathcal{S}_{\mathcal{F}}\mathcal{S}_{\mathcal{F}}&=&0\;,\;\;\textrm{if}\;\;\mathcal{S}(\mathcal{F})\;=\;0\;.
\label{22} 
\end{eqnarray}
Furthermore, the gauge fixing equation together with the Slavnov-Taylor operator, antighost equation, ghost equation and rigid operator satisfies the following algebra:
\begin{eqnarray}
\frac{\delta}{\delta b(x)}\mathcal{S}(\mathcal{F})-\mathcal{S}_{\mathcal{F}}\left(\frac{\delta\mathcal{F}}{\delta b(x)}-\partial_{\mu}A^{\mu}(x)\right)&=&\left(\frac{\delta\mathcal{F}}{\delta\bar{c}(x)}-\partial^2c(x)\right)\;,\nonumber\\
\frac{\delta}{\delta\bar{c}(x)}\mathcal{S}(\mathcal{F})+\mathcal{S}_{\mathcal{F}}\frac{\delta\mathcal{F}}{\delta\bar{c}(x)}&=&0\;,\nonumber\\
\int d^4x\;\frac{\delta}{\delta c(x)}\mathcal{S}(\mathcal{F})+\mathcal{S}_{\mathcal{F}}\int d^4x\;\left(\frac{\delta\mathcal{F}}{\delta c(x)}-\Delta_{cl}\right)&=&-i\mathcal{W}_{rig}(\mathcal{F})\;,\nonumber\\
\mathcal{W}_{rig}\mathcal{S}(\mathcal{F})-\mathcal{S}_{\mathcal{F}}\mathcal{W}_{rig}(\mathcal{F})&=&0\;.
\label{23}
\end{eqnarray}
We will see that Eqs.~\eqref{22} and \eqref{23} are very useful in the study of gauge anomalies.

\subsection{Anomalous Slavnov-Taylor operator}\label{ASTO}

One of the great questions in perturbation theory is if the symmetries of the classical theory can be implemented to quantum level, i.e., if there exist anomalies in the theory. It is worthwhile to mention that the Ward identities displayed in Eqs.~\eqref{17}-\eqref{19} are not anomalous at quantum level.\footnote{Although this is a trivial exercise, for the sake of completeness, we display the proof in Appendix \ref{QL}.} In fact, these identities allow one to eliminate a lot of renormalization parameters of the theory. Moreover, although the rigid symmetry $\mathcal{W}_{rig}$ describes the charge conservation, it does not develop an important role in the study of the quantum stability of the theory \cite{Santos:2015koa}. Indeed, the charge conservation described by the invariance of the quantum action under the operator $\mathcal{W}_{rig}$ is a global symmetry. It remains to check if the Slavnov-Taylor operator is anomalous at quantum level. In fact, this is the most important Ward identity we have because it is deeply related to gauge symmetry. To do so, let us suppose that the Slavnov-Taylor breaks down at $\hbar^n$ order in perturbation theory
\begin{eqnarray}
\mathcal{S}(\Gamma)&=&\hbar^n\Delta^{(1)}+\mathcal{O}(\hbar^{n+1})\;,
\label{24}
\end{eqnarray}
where $\Delta^{(1)}$ is a local integrated polynomial in the fields and sources with ghost number one and dimension bounded by four. Applying 
the linearized Slavnov-Taylor operator $\mathcal{S}_{\Gamma}$ in Eq.~\eqref{24}, and making use of the identity \eqref{22}, one finds
\begin{eqnarray}
\mathcal{S}_{\Gamma}\Delta^{(1)}&=&0\;.
\label{25}
\end{eqnarray}
Equation \eqref{25} is the Wess-Zumino consistency condition for the anomaly \cite{Wess:1971yu}. To solve Eq.~\eqref{25}, not that it is a cohomology problem in the space of a local integrated polynomial in the fields and their derivatives, with ghost number one and dimension bounded by four. If the cohomology of $\mathcal{S}_{\Gamma}$ is empty, we say that the model is anomaly free and the Slavnov-Taylor operator can be implemented at quantum level. On the other hand, if the cohomology of $\mathcal{S}_{\Gamma}$ is not empty, namely,
\begin{eqnarray}
\Delta^{(1)}&=&r\mathcal{A}+\mathcal{S}_{\Gamma}\hat{\Delta}^{(0)}\;,
\label{26}
\end{eqnarray}
where $\mathcal{A}\neq\mathcal{S}_{\Gamma}\hat{\mathcal{A}}$ with $\mathcal{A}$ a local integrated field polynomial and $r$ an arbitrary parameter, we have an anomaly. Thus, the Slavnov-Taylor operator only can be implemented up to $\hbar^{n-1}$ order in perturbation theory. In this case only the trivial part can be reabsorbed by redefinition of the effective action. It is worthwhile to mention that the parameter $r$ is a function of the coupling constant and cannot be determined from algebraic methods; an explicit computation of the Feynman diagrams is needed in order to determine it. However, the algebraic methods can determine the form of the functional $\mathcal{A}$ through the study of the consistency condition for the anomaly, and independently from any renormalization scheme.

Besides the restriction imposed by the Wess-Zumino consistency condition \eqref{25}, the algebra displayed at Eq.~\eqref{23} will impose a few more restrictions on the form of the anomaly -- from now on, without loss of generality \cite{Piguet:1995er}, we consider the break at $\hbar$ order. In fact, the anomaly $\Delta^{(1)}$ still must obey the following restrictions, due to \eqref{17}-\eqref{19}:
\begin{eqnarray}
\frac{\delta\Delta^{(1)}}{\delta b}&=&0\;,\nonumber\\
\frac{\delta\Delta^{(1)}}{\delta \bar{c}}&=&0\;,\nonumber\\
\int d^4x\;\frac{\delta\Delta^{(1)}}{\delta c}&=&0\;,\nonumber\\
\mathcal{W}_{rig}\Delta^{(1)}&=&0\;.
\label{27}
\end{eqnarray}
The most general solution for Eq.~\eqref{27} reads
\begin{eqnarray}
\Delta^{(1)}&=&r\int d^4x\;\partial^{\mu}c\mathcal{A}_{\mu}\;.
\label{28}
\end{eqnarray}
where $\mathcal{A}_{\mu}$ is a local functional of dimension three depending on the fundamental fields that do not belong to a BRST doublet (they belong to the trivial sector of the cohomology \cite{Piguet:1995er}), i.e., $b$, $\bar{c}$, $\lambda$, and $J$. Moreover, the dependence on the $\partial^\mu c$ is a consequence of the third equation in \eqref{27}. It is worthwhile to note that the Slavnov-Taylor operator is the functional version of the BRST operator $s$. Of course, the discrete symmetries of the action are being used to select fewer terms\footnote{We remark that the action enjoys full CPT symmetry. Furthermore, although the CPT violation is due to the background fields, the original action does not carry couplings between vector backgrounds and pseudovector operators, and vice versa.} in \eqref{28}. By eliminating gauge anomalies one understands that there must exist some mechanism yielding a vanishing parameter $r$, not that the functional $\mathcal{A}$ is not allowed. In the case that the functional $\mathcal{A}$ satisfies the Wess-Zumino consistency condition and does not belong to the trivial sector of the cohomology, an Adler-Bardeen--type nonrenormalization theorem \cite{Adler:1969er} for the gauge anomaly is needed in order to establish the behavior of the parameter $r$ under quantum corrections. For instance, whether the parameter $r$ vanishes at one-loop order this theorem says if this property is kept to all orders, assuring (or not) the absence of gauge anomaly to all orders in perturbation theory.

Let us see in the sequel the possible contributions for the gauge anomaly respecting the criteria above mentioned and the conditions displayed at Eqs.~\eqref{25} and \eqref{28}. We have
\begin{align}
\Delta^{(1)}&=\int d^4x\;\left(r_1D^{\alpha}_{\phantom{\alpha}\mu}\epsilon_{\nu\alpha\rho\sigma}\partial^{\mu}cA^{\nu}\partial^{\rho}A^{\sigma}+r_2D^{\alpha}_{\phantom{\alpha}\nu}\epsilon_{\mu\alpha\rho\sigma}\partial^{\mu}cA^{\nu}\partial^{\rho}A^{\sigma}+r_3D^{\beta\alpha}\epsilon_{\beta\alpha\mu\nu}\partial^{\mu}cA^{\sigma}\partial_{\sigma}A^{\nu}\right.\displaybreak[3]\nonumber\\
&+\left.r_4D^{\beta\alpha}\epsilon_{\beta\alpha\mu\nu}\partial_{\sigma}cA^{\mu}\partial^{\sigma}A^{\nu}+r_5D^{\beta\alpha}\epsilon_{\beta\alpha\mu\nu}\partial^{\mu}cA_{\sigma}\partial^{\nu}A^{\sigma}+r_6D^{\beta\alpha}\epsilon_{\beta\alpha\mu\nu}\partial_{\sigma}cA^{\mu}\partial^{\nu}A^{\sigma}\right.\displaybreak[3]\nonumber\\
&+\left.r_7D^{\beta\alpha}\epsilon_{\beta\alpha\mu\nu}\partial_{\sigma}cA^{\sigma}\partial^{\mu}A^{\nu}+r_{8}D^{\lambda}_{\phantom{\lambda}\mu}\epsilon_{\lambda\nu\rho\sigma}\partial^{\rho}cA^{\sigma}\partial^{\mu}A^{\nu}+r_{9}D^{\kappa\lambda}\epsilon_{\kappa\lambda\mu\nu}\partial^{\mu}cA^{\nu}\partial_{\rho}A^{\rho}\right.\displaybreak[3]\nonumber\\
&+\left.r_{10}\partial^{\mu}c\bar{A}_{\mu}A_{\alpha}A^{\alpha}+r_{11}\partial^{\mu}cA_{\mu}\bar{A}_{\alpha}A^{\alpha}+
r_{12}C^{\mu\nu}\partial_{\mu}cA_{\nu}A_{\alpha}A^{\alpha}\right.\displaybreak[3]\nonumber\\
&+\left.r_{13}C^{\mu\nu}\partial_{\nu}cA_{\mu}A_{\alpha}A^{\alpha}+r_{14}C^{\mu\nu}A_{\mu}A_{\nu}A^{\alpha}\partial_{\alpha}c+r_{15}C^{\mu\nu}\partial_{\mu}c\overline{\psi}\gamma_{\nu}\psi+r_{16}D^{\mu\nu}\partial_{\mu}c\overline{\psi}\gamma_5\gamma_{\nu}\psi\right.\displaybreak[3]\nonumber\\
&+\left.r_{17}E^{\mu}\partial_{\mu}c\overline{\psi}\psi+r_{18}F^{\mu}\partial_{\mu}c\overline{\psi}\gamma_5\psi+r_{19}G^{\alpha\beta\mu}\partial_{\mu}c\overline{\psi}\sigma_{\alpha\beta}\psi+r_{20}\epsilon_{\mu\nu\alpha\beta}\partial^{\mu}cB^{\nu}\partial^{\alpha}A^{\beta}\right.\displaybreak[3]\nonumber\\
&+\left.r_{21}\partial^{\mu}cA_{\mu}A_{\alpha}A^{\alpha}+r_{22}\partial^{\mu}c\overline{\psi}\gamma_{\mu}\psi+r_{23}\partial^{\mu}c\partial^{\nu}F_{\mu\nu}\right)\;.
\label{29}
\end{align}
Since $\Delta^{(1)}$ must satisfy the consistency condition \eqref{25}, we get the following relations between some parameters:
\begin{eqnarray}
r_1=r_2\;,\;\;\;r_3=r_4\;,\;\;\;r_5=r_6\;,\;\;\;r_8=r_9=0\;,\;\;\;r_{10}=\frac{r_{11}}{2}\;,\;\;\;r_{12}=r_{13}=\frac{r_{14}}{2}\;.
\label{30}
\end{eqnarray}
Thus, Eq.~\eqref{29} becomes
\begin{align}
\Delta^{(1)}&=r_1\int d^4x\left\{\left[(\partial^{\mu}D^{\alpha}_{\phantom{\alpha}\mu})\epsilon_{\alpha\nu\rho\sigma}-(\partial^{\mu}D^{\alpha}_{\phantom{\alpha}\nu})\epsilon_{\mu\alpha\rho\sigma}\right]cA^{\nu}\partial^{\rho}A^{\sigma}\right.\displaybreak[3]\nonumber\\
&+\left.\frac{1}{2}\left(D^{\alpha}_{\phantom{\alpha}\mu}\epsilon_{\alpha\nu\rho\sigma}-D^{\alpha}_{\phantom{\alpha}\nu}\epsilon_{\mu\alpha\rho\sigma}+D^{\alpha}_{\phantom{\alpha}\rho}\epsilon_{\mu\nu\alpha\sigma}-D^{\alpha}_{\phantom{\alpha}\sigma}\epsilon_{\mu\nu\rho\alpha}\right)c\partial^{\mu}A^{\nu}\partial^{\rho}A^{\sigma}\right.\displaybreak[3]\nonumber\\
&+\left.D^{\alpha}_{\phantom{\alpha}\rho}\epsilon_{\sigma\alpha\mu\nu}cA^{\mu}\partial^{\rho}\partial^{\sigma}A^{\nu}\right\}\nonumber\\
&-\int d^4x\left\{r_3\left[(\partial^{\mu}D^{\beta\alpha})\epsilon_{\beta\alpha\mu\nu}cA_{\sigma}\partial^{\sigma}A^{\nu}+(\partial^{\sigma}D^{\beta\alpha})\epsilon_{\beta\alpha\mu\nu}cA^{\mu}\partial_{\sigma}A^{\nu}\right]\right.\displaybreak[3]\nonumber\\
&+\left.r_5\left[(\partial^{\mu}D^{\beta\alpha})\epsilon_{\beta\alpha\mu\nu}cA_{\sigma}\partial^{\nu}A^{\sigma}+(\partial^{\sigma}D^{\beta\alpha})\epsilon_{\beta\alpha\mu\nu}cA^{\mu}\partial^{\nu}A_{\sigma}\right]+r_7(\partial^{\sigma}D^{\beta\alpha})\epsilon_{\beta\alpha\mu\nu}cA_{\sigma}\partial^{\mu}A^{\nu}\right.\displaybreak[3]\nonumber\\
&+\left.\frac{1}{2}D^{\beta\alpha}\left[r_7(\epsilon_{\beta\alpha\mu\nu}\eta_{\rho\sigma}+\epsilon_{\beta\alpha\rho\sigma}\eta_{\mu\nu})+(r_3-r_5)(\epsilon_{\beta\alpha\mu\sigma}\eta_{\rho\nu}+\epsilon_{\beta\alpha\rho\nu}\eta_{\sigma\mu})\right]c\partial^{\mu}A^{\nu}\partial^{\rho}A^{\sigma}\right.\displaybreak[3]\nonumber\\
&+\left.D^{\beta\alpha}\epsilon_{\beta\alpha\mu\nu}c\left[r_3A^{\mu}\partial^2A^{\nu}+r_5A^{\mu}\partial^{\nu}\partial_{\sigma}A^{\sigma}+(r_3+r_7)A_{\sigma}\partial^{\sigma}\partial^{\mu}A^{\nu}\right]\right\}\displaybreak[3]\nonumber\\
&+\int d^4x\left[r_{10}\left(\partial^{\mu}c\bar{A}_{\mu}A_{\alpha}A^{\alpha}+2\partial^{\mu}cA_{\mu}\bar{A}_{\alpha}A^{\alpha}\right)+
r_{12}\left(C^{\mu\nu}\partial_{\mu}cA_{\nu}A_{\alpha}A^{\alpha}\right.\right.\displaybreak[3]\nonumber\\
&+\left.\left.C^{\mu\nu}\partial_{\nu}cA_{\mu}A_{\alpha}A^{\alpha}+2C^{\mu\nu}A_{\mu}A_{\nu}A^{\alpha}\partial_{\alpha}c\right)+r_{15}C^{\mu\nu}\partial_{\mu}c\overline{\psi}\gamma_{\nu}\psi+r_{16}D^{\mu\nu}\partial_{\mu}c\overline{\psi}\gamma_5\gamma_{\nu}\psi\right.\displaybreak[3]\nonumber\\
&+\left.r_{17}E^{\mu}\partial_{\mu}c\overline{\psi}\psi+r_{18}F^{\mu}\partial_{\mu}c\overline{\psi}\gamma_5\psi+r_{19}G^{\alpha\beta\mu}\partial_{\mu}c\overline{\psi}\sigma_{\alpha\beta}\psi+r_{20}\epsilon_{\mu\nu\alpha\beta}\partial^{\mu}cB^{\nu}\partial^{\alpha}A^{\beta}\right.\displaybreak[3]\nonumber\\
&+\left.r_{21}\partial^{\mu}cA_{\mu}A_{\alpha}A^{\alpha}+r_{22}\partial^{\mu}c\overline{\psi}\gamma_{\mu}\psi+r_{23}\partial^{\mu}c\partial^{\nu}F_{\mu\nu}\right]\;.
\label{31}
\end{align}
However, it is possible to show that $\Delta^{(1)}=\mathcal{S}_{\Sigma}\hat{\Delta}^{(0)}$, where
\begin{eqnarray}
\hat{\Delta}^{(0)}&=&\frac{r_1}{2}\int d^4x\left(D^{\alpha}_{\phantom{\alpha}\mu}\epsilon_{\alpha\nu\rho\sigma}-D^{\alpha}_{\phantom{\alpha}\nu}\epsilon_{\mu\alpha\rho\sigma}\right)A^{\mu}A^{\nu}\partial^{\rho}A^{\sigma}\nonumber\\
&-&\int d^4x\;D^{\beta\alpha}\epsilon_{\beta\alpha\mu\nu}\left(r_3A^{\mu}A_{\sigma}\partial^{\sigma}A^{\nu}+r_5A^{\mu}A_{\sigma}\partial^{\nu}A^{\sigma}+\frac{r_7}{2}A_{\alpha}A^{\alpha}\partial^{\mu}A^{\nu}\right)\nonumber\\
&-&\int d^4x\left[r_{10}A^{\mu}\bar{A}_{\mu}A_{\alpha}A^{\alpha}+
r_{12}C^{\mu\nu}A_{\mu}A_{\nu}A_{\alpha}A^{\alpha}+r_{15}C^{\mu\nu}A_{\mu}\overline{\psi}\gamma_{\nu}\psi+r_{16}D^{\mu\nu}A_{\mu}\overline{\psi}\gamma_5\gamma_{\nu}\psi\right.\nonumber\\
&+&\left.r_{17}E^{\mu}A_{\mu}\overline{\psi}\psi+r_{18}F^{\mu}A_{\mu}\overline{\psi}\gamma_5\psi+r_{19}G^{\alpha\beta\mu}A_{\mu}\overline{\psi}\sigma_{\alpha\beta}\psi+r_{20}\epsilon_{\mu\nu\alpha\beta}B^{\nu}A^{\mu}\partial^{\alpha}A^{\beta}\right.\nonumber\\
&+&\left.\frac{r_{21}}{4}A_{\mu}A^{\mu}A_{\alpha}A^{\alpha}+r_{22}A^{\mu}\overline{\psi}\gamma_{\mu}\psi\right]\;.
\label{32}
\end{eqnarray}
Thus, $\Delta^{(1)}$ belongs to the trivial sector of cohomology and does not contribute to the anomaly. Since this term belongs to the trivial sector of the cohomology it is an ambiguous term and can be compensated by introducing suitable noninvariant counterterm $-\hat{\Delta}^{(0)}$ into the effective action. Note, from Eq.~\eqref{31}, that the composite operator $c\partial^{\mu}A^{\nu}\tilde{F}_{\alpha\nu}$, which couples to the $D^{\alpha}_{\phantom{\lambda}\mu}$ source, has the same discrete symmetries of the composite operator $\overline{\psi}\gamma_{5}\gamma_{\nu}D_{\mu}\psi$. For this reason the $C^{\mu\nu}$ source cannot couple to it. Otherwise, the sources $C^{\mu\nu}$ and $D^{\mu\nu}$ would be indistinguishable at quantum level.

\section{Conclusion}\label{FINAL}

Applying the BRST quantization together with the algebraic renormalization technique, we conclude that the solution of the cohomology of the Slavnov-Taylor operator in the space of the local integrated polynomial of dimension four and ghost number one is empty in the considered order. Thus, the existent solutions belong to the trivial sector of the cohomology, and a non-nonrenormalization theorem is need. Since the method here employed is recursive, we concluded that the Lorentz-violating QED is free of gauge anomalies to all orders in perturbation theory. 

As aforementioned, the quantum stability of the theory is analyzed by seeking the most general solutions of the Slavnov-Taylor operator in the space of the local integrated polynomials of dimension four and ghost number zero, i.e., $\mathcal{S}_{\Gamma}\Gamma=0$, where $\Gamma=\Gamma_0+\mathcal{S}_{\Gamma}\Delta^{(-1)}$, and $\Gamma_0$ correspond to the nontrivial solutions. This means seeking BRST invariant solutions for the effective action. This was already performed at Ref.~\cite{Santos:2015koa}, showing the renormalizability of the Lorentz-violating QED to all orders.

It is worth mentioning that the model presented here does not enjoy chiral symmetry, not even classically. The chiral symmetry is softly broken just as in the usual massive QED. However, the usual partially conserved axial current (PCAC) \cite{Bell:1969ts} could be modified already at the classical level by terms proportional to the background fields. Moreover, since the chiral current does not directly interact with the gauge field (one of the consequences for the model to be gauge anomaly free), the contribution to the chiral anomaly could only come from the usual ABJ chiral anomaly \cite{Bell:1969ts,Adler:1969gk} and terms proportional to the background fields. Thus, the decay of the pion would possibly be corrected by terms proportional to the background fields. Nevertheless, the issue of the chiral anomaly is outside the scope of this work and is left for future investigation.

\appendix

\section{Quantum consistency of the gauge fixing, antighost and ghost equations}\label{QL}

Let us show that the Ward identities displayed at Eqs.~\eqref{17}-\eqref{20} hold at quantum level. First, we will see the gauge fixing equation, whose operator obeys
\begin{eqnarray}
\left[\frac{\delta}{\delta b(x)},\frac{\delta}{\delta b(y)}\right]&=&0\;.
\label{AP1}
\end{eqnarray}
Let $\Gamma$ be a quantum action, i.e.,
\begin{eqnarray}
\Gamma&=&\sum_{n=0}^{\infty}\hbar^n\Gamma^{(n)}\;\;\textrm{where}\;\;\Gamma^{(0)}\;=\;\Sigma\;.
\label{AP2}
\end{eqnarray}
From the QAP we can suppose that the gauge fixing equation holds up to $\hbar^{n-1}$ order in perturbation theory, namely,
\begin{eqnarray}
\frac{\delta \Gamma}{\delta b(x)}&=&\partial_{\mu}A^{\mu}(x)+\Delta\cdot\Gamma\;=\;\partial_{\mu}A^{\mu}(x)+\hbar^n\Delta(x)+\mathcal{O}(\hbar^{n+1})\;,
\label{AP3}
\end{eqnarray}
where $\Delta(x)$ is a local polynomial insertion of dimension two and ghost number zero depending on the fields, their derivatives, and the sources. The most general form for it is
\begin{eqnarray}
\Delta(x)&=&F(A,\bar{c},c,\mathcal{J})(x)+\alpha b(x)\;,
\label{AP4}
\end{eqnarray}
where $F$ is a local polynomial in the fields $A, \bar{c}, c$ and sources $\mathcal{J}$, and $\alpha$ is a constant parameter. From Eq.~\eqref{AP1} we get the following consistency condition:
\begin{eqnarray}
\frac{\delta}{\delta b(y)}\Delta(x)-\frac{\delta}{\delta b(x)}\Delta(y)&=&0\;.
\label{AP5}
\end{eqnarray}
From Eq.~\eqref{AP4} we get
\begin{eqnarray}
\frac{\delta}{\delta b(y)}\Delta(x)&=&\frac{\delta}{\delta b(x)}(F(A,\bar{c},c,\mathcal{J})(y)+\alpha b(y))\;.
\label{AP6}
\end{eqnarray}
Integrating out this equation, we get
\begin{eqnarray}
\Delta(x)&=&\frac{\delta}{\delta b(x)}\int d^4y\;\left(F(A,\bar{c},c,\mathcal{J})(y)b(y)+\frac{\alpha}{2} b(y)b(y)\right)\;.
\label{AP7}
\end{eqnarray}
Thus the gauge fixing equation \eqref{AP3} becomes
\begin{eqnarray}
\frac{\delta \bar{\Gamma}}{\delta b(x)}&=&\partial_{\mu}A^{\mu}(x)+\mathcal{O}(\hbar^{n+1})\;,
\label{AP8}
\end{eqnarray}
where we have redefined the quantum action as
\begin{eqnarray}
\bar{\Gamma}&=&\Gamma-\hbar^n\int d^4y\;\left(F(A,\bar{c},c,\mathcal{J})(y)b(y)+\frac{\alpha}{2} b(y)b(y)\right)\;.
\end{eqnarray}
We can repeat this procedure recursively, and we shown that the gauge fixing equation holds to all orders. 

Let us now investigate the quantum breaking of the antighost equation. Again, employing the QAP,
\begin{eqnarray}
\frac{\delta \bar{\Gamma}}{\delta \bar{c}(x)}&=&\partial^2c(x)+\Delta\cdot\bar{\Gamma}\;=\;\partial^2c(x)+\hbar^n\Delta(x)+\mathcal{O}(\hbar^{n+1})\;,
\label{AP9}
\end{eqnarray}
with $\Delta(x)$ being a local polynomial of dimension two and ghost number $+1$. Its general form reads
\begin{eqnarray}
\Delta(x)&=&G(A,c,\mathcal{J})(x)+f(c)\bar{c}(x)\;.
\label{AP10}
\end{eqnarray}
Here, $G(A,c,\mathcal{J})(x)$ does not depend on the $b$ field, since its dependence was already absorbed into the effective action $\bar{\Gamma}$. The function $f(c)$ is supposed to depend on $c$, but, from Table \ref{table1}, it is clear that it is impossible to build a function $f(c)$ with mass dimension two and ghost number two depending only on $c$. Then, we set $f(c)=0$. From the algebra obeyed by the antighost
\begin{eqnarray}
\left\{\frac{\delta}{\delta \bar{c}(x)},\frac{\delta}{\delta  \bar{c}(y)}\right\}&=&0\;,
\label{AP11}
\end{eqnarray}
we derive another consistency condition
\begin{eqnarray}
\frac{\delta}{\delta\bar{c}(y)}\Delta(x)+\frac{\delta}{\delta\bar{c}(x)}\Delta(y)&=&0\;.
\label{AP12}
\end{eqnarray}
The solution for \eqref{AP12} reads
\begin{eqnarray}
\Delta(x)&=&\frac{\delta}{\delta \bar{c}(x)}\int d^4y\;\bar{c}(y)G(A,c,\mathcal{J})(y)\;.
\label{AP13}
\end{eqnarray}
Rewriting the antighost equation
\begin{eqnarray}
\frac{\delta \tilde{\Gamma}}{\delta \bar{c}(x)}&=&\partial^2c(x)+\mathcal{O}(\hbar^{n+1})\;,
\label{AP14}
\end{eqnarray}
with
\begin{eqnarray}
\tilde{\Gamma}&=&\bar{\Gamma}-\hbar^n\int d^4y\;\bar{c}(y)G(A,c,\mathcal{J})(y)\;
\label{AP15}
\end{eqnarray}
being the redefined effective action. Again, we see that the antighost equation holds at considered order. 

Finally, supposing that the ghost equation \eqref{18} breaks down at $\hbar^n$ order, i.e.,
\begin{eqnarray}
\frac{\delta \tilde{\Gamma}}{\delta c(x)}&=&\Delta_{cl}+\Delta\cdot\tilde{\Gamma}\;=\;\Delta_{cl}+\hbar^n\Delta(x)+\mathcal{O}(\hbar^{n+1})\;,
\label{AP16}
\end{eqnarray}
where $\Delta_{cl}$ stands for the classical breaking and $\Delta(x)$ is a local polynomial of dimension four and ghost number $-1$. Its most general form reads
\begin{eqnarray}
\Delta(x)&=&\int d^4x\left(a_1\overline{Y}\psi+a_2\overline{\psi}Y+L(A,\mathcal{J},a_{i+2})(x)\right)\;,
\label{AP17}
\end{eqnarray}
where the $a$'s are arbitrary dimensionless parameters,\footnote{The parameters $a_{i+2}$ (with $i=1,2,\dots$) indicate one different parameter for each different combination between sources and the gauge field.} and $L$ is a local polynomial depending on the field $A$ and sources $\mathcal{J}$. Note that the antighost field cannot enter into $L$ since its contribution to the quantum level was already absorbed into $\tilde{\Gamma}$ as well as $b$. Thus, it is clear that we can write 
\begin{eqnarray}
\Delta(x)&=&\frac{\delta}{\delta c(x)}\int d^4y\;c(y)\left(a_1\overline{Y}\psi+a_2\overline{\psi}Y+L(A,\mathcal{J},a_{i+2})(y)\right)\;,
\label{AP18}
\end{eqnarray}
and
\begin{eqnarray}
\frac{\delta \hat\Gamma}{\delta c(x)}&=&\Delta_{cl}+\mathcal{O}(\hbar^{n+1})\;,
\label{AP19}
\end{eqnarray}
with
\begin{eqnarray}
\hat\Gamma&=&\tilde{\Gamma}-\hbar^n\int d^4y\;c(y)\left(a_1\overline{Y}\psi+a_2\overline{\psi}Y+L(A,\mathcal{J},a_{i+2})(y)\right)\;.
\label{AP20}
\end{eqnarray}
We have shown that the ghost equation holds at $\hbar^n$ order. The proof to all orders follows trivially by induction. The proof that the rigid symmetry holds at $\hbar^n$ order follows by supposing a general functional respecting the power-counting criteria, being a Lorentz scalar and a scalar in the spinor space, with vanishing ghost number.

\section*{Acknowledgements}

The Conselho Nacional de Desenvolvimento Cient\'{i}fico e Tecnol\'{o}gico (CNPq-Brazil), The Coordena\c c\~ao de Aperfei\c coamento de Pessoal de N\'ivel Superior (CAPES), and the Pr\'o-Reitoria de Pesquisa, P\'os-Gradua\c c\~ao e Inova\c c\~ao (PROPPI-UFF) are acknowledged for financial support.

\end{document}